\documentclass[%
reprint,
groupedaddress,
nofootinbib,
 amsmath,amssymb,
 aps,
prl,
floatfix,
]{revtex4-2}

\usepackage{graphicx}	
\usepackage{bm}	
\usepackage{amssymb}
\usepackage{multirow}
\usepackage{array}
\usepackage{color,soul}
\usepackage{float}
\usepackage{tabularx}
\usepackage{xr}
\usepackage{upgreek}

\begin{document}
\title{Dynamics of spin glasses in two dimensions}
\author{Hongze~Li}
\author{Raymond~L.~Orbach}
\email[Contact author: ]{orbach@utexas.edu}
\affiliation{Texas Materials Institute, The University of Texas at Austin, Austin, Texas  78712, USA}
\author{Gregory G. Kenning}
\affiliation{Madia Department of Chemistry, Biochemistry, Physics and Engineering, Indiana 
University of Pennsylvania, Indiana, Pennsylvania 15705, USA}

\date{\today}   

\begin{abstract}
\noindent Spin glass dynamics is a strong function of spatial dimensionality $D$.  The lower critical dimension is close to 2.5, so that, in two dimensions, the condensation temperature $T_\text{g}=0$, and only fluctuations are present at finite temperatures.  However, by using thin film multilayers, one can explore the dynamics in both $D=3$ and $D=2$ dimensions. Spin glass thin film multilayers transition from $D=3$ dynamics at short to intermediate times to $D = 2$ dynamics at long times. Correlation lengths of CuMn 4.5 nm multilayers at long times are shown to be grow more rapidly in $D=2$ as compared to $D=3$, and for the longest measurement time, experimentally reach equilibrium in qualitative agreement with simulations. 
\end{abstract}
\pacs{}

\maketitle

\maketitle
\noindent Spin glass dynamics are extraordinarily sensitive to spatial dimensions.  For Ising spin glasses, a finite transition temperature $T_\text{g}$ exists for $D = 3$, while $T_\text{g} = 0$ K in $D = 2$.  This is a consequence of the lower critical dimension $D\sim2.5$ lying between $D = 3$ and $D = 2$ \cite{Bray:86,Franz:94,Boettcher:05,Maiorano:18}.  As a consequence, the growth of the coherence length in bulk samples (i.e. $D = 3$) has been extracted from theory, simulations, and experiments \cite{Dahlberg:25}.  However, the properties of the correlation length for $D = 2$ spin glasses are only known experimentally in the limit $T \rightarrow 0$ \cite{Dekker:88a,Dekker:88b,Dekker:88c,Dekker:89} because only fluctuations are present for $T>T_\text{g}=0$.  However, multilayer spin glasses provide a stable $D = 2$ configuration, allowing experimental measurement of the correlation length dynamics at finite temperatures.\\
\\
This paper reports $D = 2$ dynamics utilizing a multilayer sample of thin CuMn layers separated by thick Cu layers. Multilayers are used to enhance the total magnetic moment to a reasonably high level. Their overall dynamics have been analyzed previously, both experimentally \cite{Guchhait:15,Kenning:90,Zhai:17a,Zhai:17b,Zhai:24} and theoretically \cite{Fernandez:19a} in terms of an initial $D = 3$ growth from nucleation for short times, ``crossing over'' at a time $t_{\text {co}}$ to $D = 2$ when the spin glass coherence length becomes comparable to the CuMn layer thickness.  The correlated region has been described as ``pancake-like'' \cite{Zhai:17b} in that the thickness is constrained by the dimensions of the CuMn thin film (typically in the 4.5 to 20 nm range), but allowed to grow in the transverse direction, i.e. in $D = 2$.  Fig. 1 represents a pictorial view of the time evolution of the correlated state.\\
\\
By making measurements at longer times than previous \cite{Zhai:17a,Zhai:17b}, and with a judicious selection of measurement temperatures, we are able to track the growth of the correlation length in $D = 2$ from $t_{\text {co}}$ to its equilibrium value $\xi_{\text {eq}}(T) \sim a_0 (T_\text{g}/T)^{\nu_{2d}},$ where $a_0$ is of the order of the average separation of Mn ions, and $\nu_{2d}$ is the critical exponent for the correlation length in $D = 2$.
\begin{figure}[htbp]
    \centering
    \includegraphics[width=0.49\textwidth]{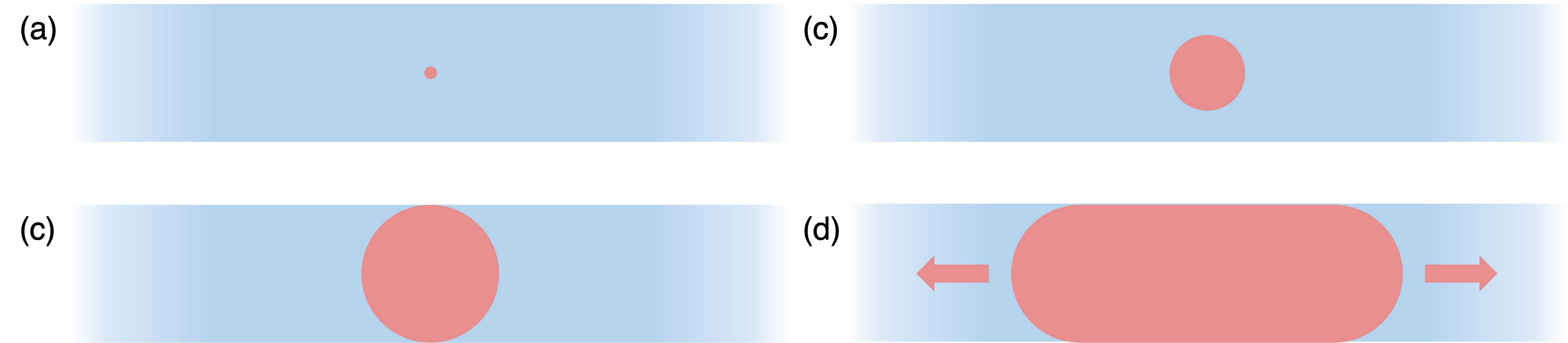}
    \caption{(a) Nucleation at $t = 0$. (b) Growth as in $D=3$ before crossover ($t < t_\text{co}$). (c) Growth hits the thin film thickness  at crossover ($t = t_\text{co}$). (d) Growth in the parallel direction after crossover ($t > t_\text{co}$) with dynamics in $D=2$.}
    \label{fig:schematic}
\end{figure}

\noindent The correlated volume for a thin film of thickness $\mathcal L$ after crossover is described in detail in \cite{Zhai:17b}.  Upon reaching equilibrium, the correlation lengths perpendicular and parallel to the film are given by \cite{Young:14,Victor:16},
\begin{equation}
\begin{split}
    \xi_{\perp}(t_{\text {co}},T)={\mathcal L}\\
    \xi_{\parallel} = b\,\bigg({\frac {T_\text{g}}{T}}\bigg)^{\nu_{2d}}{\mathcal L}
\end{split}
\end{equation}
where $b$ is a constant of order unity. The correlated volume is,
\begin{equation}
    V_{\text {corr}} = \pi\,b^2\,{\mathcal L}^3\bigg({\frac{T_\text{g}}{T}}\bigg)^{2\nu_{2d}}.
\end{equation}
The fitted variable constant $b$ is found to be 1.6, 3.45, or 2.46 for Ising, Heisenberg spins, or chiral correlation lengths, respectively \cite{Zhai:17b}.\\
\\
We adopt the notation of the simulations \cite{Baity:18} for the temperature and time dependence of the coherence length (before crossover) and the correlation length (after crossover):
\begin{equation}
\xi(t_w,T) = a_0c_1\,\bigg({\frac {t_w}{\tau_0}}\bigg)^{T/(T_\text{g}Z_c)}
\end{equation}
where $t_w$ is the aging (waiting) time, $\tau_0$ is an exchange time, $a_0$ is the average distance between Mn ions, and $c_1$ a constant of order unity.
The rate of growth is then set by the value of $Z_c$ which itself is temperature dependent in $D = 3$. See for example Fig. 4 of \cite{Baity:18} where $Z_c$ is plotted against $T/T_\text{g}$, rising from a value of 6.69(6) at $T=T_\text{g}$ to values as large as 10\,-\,15 at $T=0.5\,T_\text{g}$ for $T\sim T_\text{g}/2$ depending on the model dynamics. When the coherence length is less than $\mathcal L$ (i.e. in the $D = 3$ time domain before crossover) $Z_c\sim 9.6$ \cite{Zhai:17a}.  However, for D = 2 spin glasses, a scaling study by Fernandez et al. \cite{Fernandez:19b} finds,
\begin{equation}
\xi(t_w) \sim a_0\,\bigg({\frac {t_w}{\tau_0}}\bigg)^{1/Z_c},
\end{equation}
{\it independent of temperature $T$} with $Z_c \approx7.14$.  This implies that the correlation length grows more rapidly in D = 2 than the coherence length in $D = 3$ because $Z_c$ is much larger than 7 in D = 3 \cite{Baity:18}.\\
\begin{figure}[htbp]
    \centering
    \includegraphics[width=0.49\textwidth]{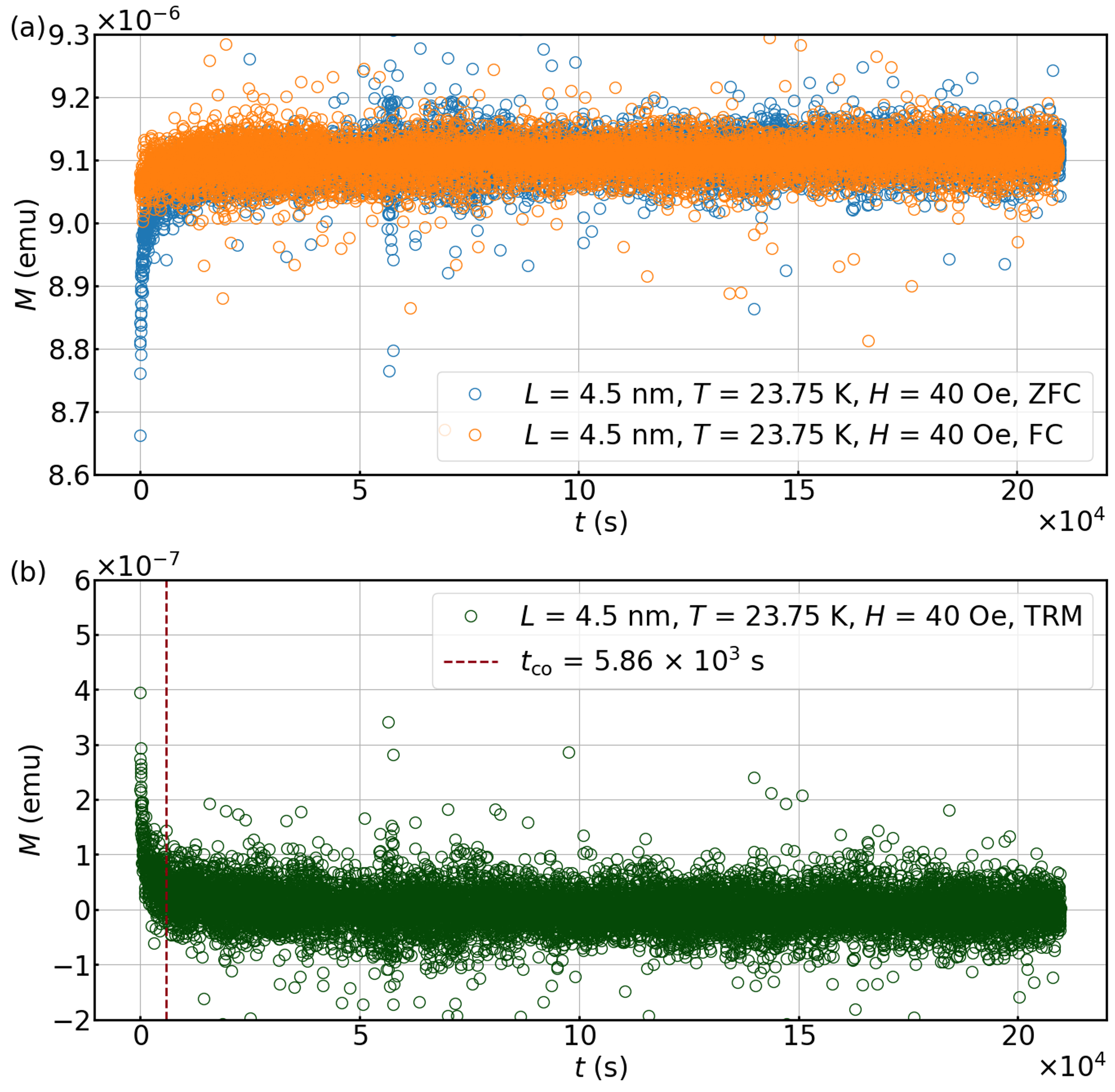}
    \caption{(a) Typical $M_\text{ZFC}(t)$ and $M_\text{FC}(t)$ at $T=23.75$ K and $H=40$ Oe. (b) Typical $M_\text{TRM}(t)$ at $T=23.75$ K and $H=40$ Oe, obtained through $M_\text{TRM}(t)=\alpha_f M_\text{FC}(t)-M_\text{ZFC}(t)$ with $\alpha_f=1.000$. The calculated $t_\text{co}=5.83\times10^3$ s at $T=23.75$ K is labeled in panel (b).}
    \label{fig:zfc_fc_trm}
\end{figure}

\noindent CuMn/Cu multilayer samples consisting of 40 bilayers (4.5 nm of CuMn and 60 nm of Cu) were DC sputtered from a 99.999\% CuMn target with a nominal Mn concentration of 13.5 at.\% and a 99.999\% Cu target. $T_\text{g}$ obtained from ``bulk'' CuMn/Cu multilayer samples with the CuMn layer thickness of $\sim 1\,\mu$m translates to a Mn concentration of 11.7 at.\% \cite{PhysRevLett.54.150,PhysRevB.33.4803}. The freezing temperature $T_\text{f}$ of the CuMn/Cu multilayer samples is determined by the onset of irreversibility from field-cooled (FC) and
zero-field-cooled (ZFC) magnetization measurements: $T_\text{f}\approx24$ K for the 4.5 nm sample. Our measurement temperatures are set below $T_\text{f}$.\\
\\
Our protocol measures the time dependence of the ZFC and FC magnetizations, $M_\text{ZFC}(t, T)$ and $M_\text{FC}(t, T)$, at a time scale longer than $2\times10^5$ s as displayed in Fig. \ref{fig:zfc_fc_trm}(a). We can extract the thermoremanent magnetization, $M_\text{TRM}(t, T)$, from the extended principle of superposition \cite{doi:10.1142/9789812819437_0001}, as shown in Fig. \ref{fig:zfc_fc_trm}(b):
\begin{equation}
    M_\text{ZFC}(t, T)+M_\text{TRM}(t, T)=M_\text{FC}(t, T)
\end{equation}
Following \cite{Zhai:17a}, we also introduce a scaling factor $\alpha=M_\text{ZFC}(t, T)/M_\text{FC}(t, T)$ and its limit $\alpha_f$ at long time scale, typically at $t>2\times10^5$ s, to adjust $M_\text{TRM}(t, T)=\alpha_f M_\text{FC}(t, T)-M_\text{ZFC}(t, T)$ when equilibrium is reached (see our later discussion). The adjustments using $\alpha_f$ are subtle, with $\epsilon=1-\alpha_f<0.003$ in our measurements. \\
\\
We extract a crossover time, $t_{\text {co}}$, where $D=3$ dynamics change to $D=2$ dynamics, as shown in Fig. \ref{fig:schematic}(c), from
\begin{equation}
    \xi(t_\text{co},T)=a_0c_1(t_\text{co}/\tau_0)^{T/(T_\text{g}Z_c)}={\mathcal L}.
    \label{eq:t_co_cal}
\end{equation}
Here, $a_0=0.523$ nm, $c_1\approx1.448$, and $Z_c\approx9.62$ from \cite{Zhai:17a}. For example, at $T_\text{m}=23$ K, $t_\text{co}\approx2.04\times10^4$ s; at $T_\text{m}=23.75$ K, $t_\text{co}\approx5.86\times10^3$ s.\\
\begin{figure}[htbp]
    \centering
    \includegraphics[width=0.49\textwidth]{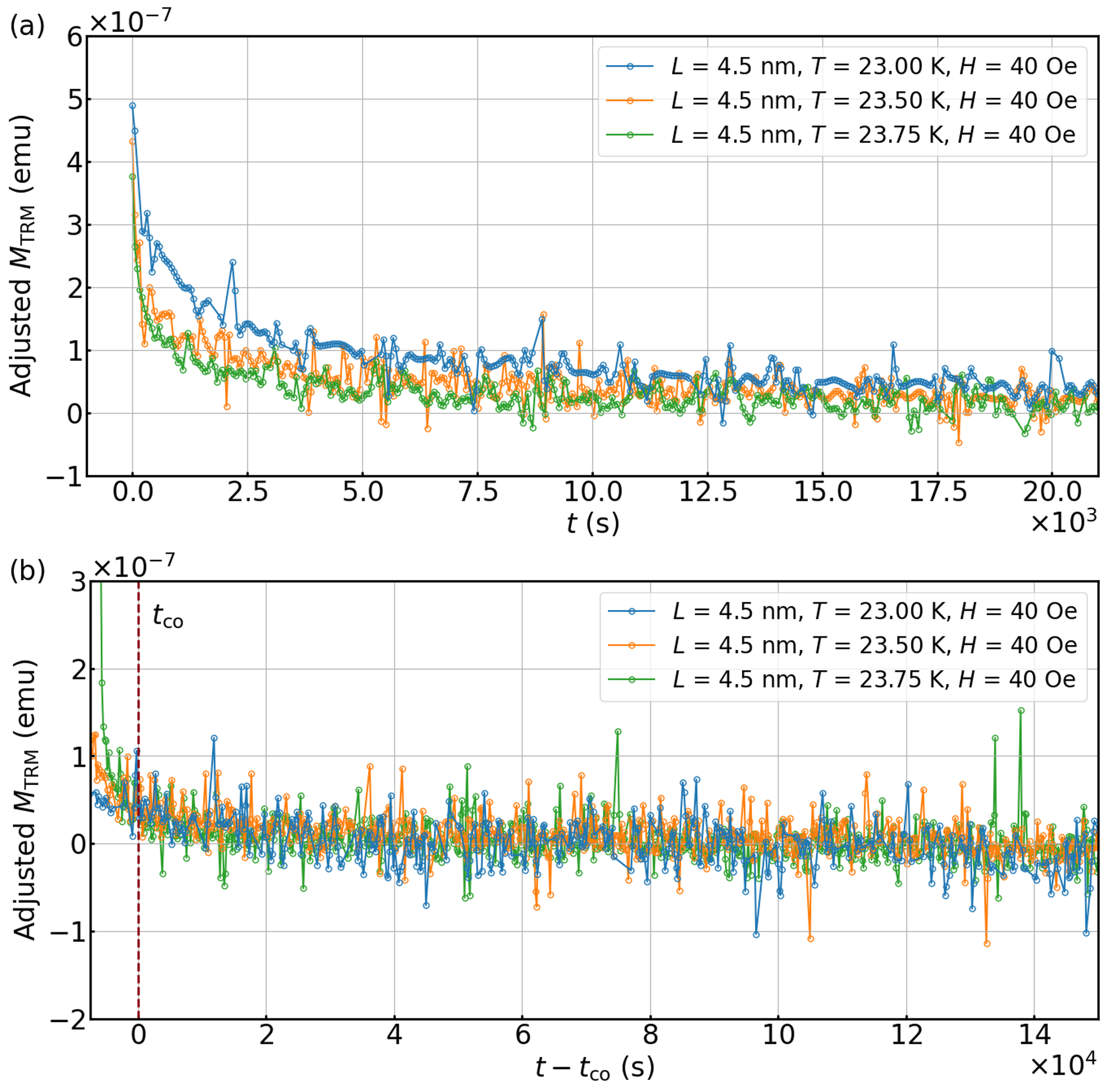}
    \caption{$\alpha_f$-adjusted (a) $M_\text{TRM}(t,T)$ at short time scales and (b) $M_\text{TRM}(t-t_\text{co},T)$ at long time scales for different measurement temperatures. The data size is reduced to (a) 4000 points and (b) 1000 points over the entire measurement time using cubic spline interpolation to better demonstrate the differences as a function of temperature.}
    \label{fig:2d_3d_dynamics}
\end{figure}

\begin{figure}[htbp]
    \centering
    \includegraphics[width=0.49\textwidth]{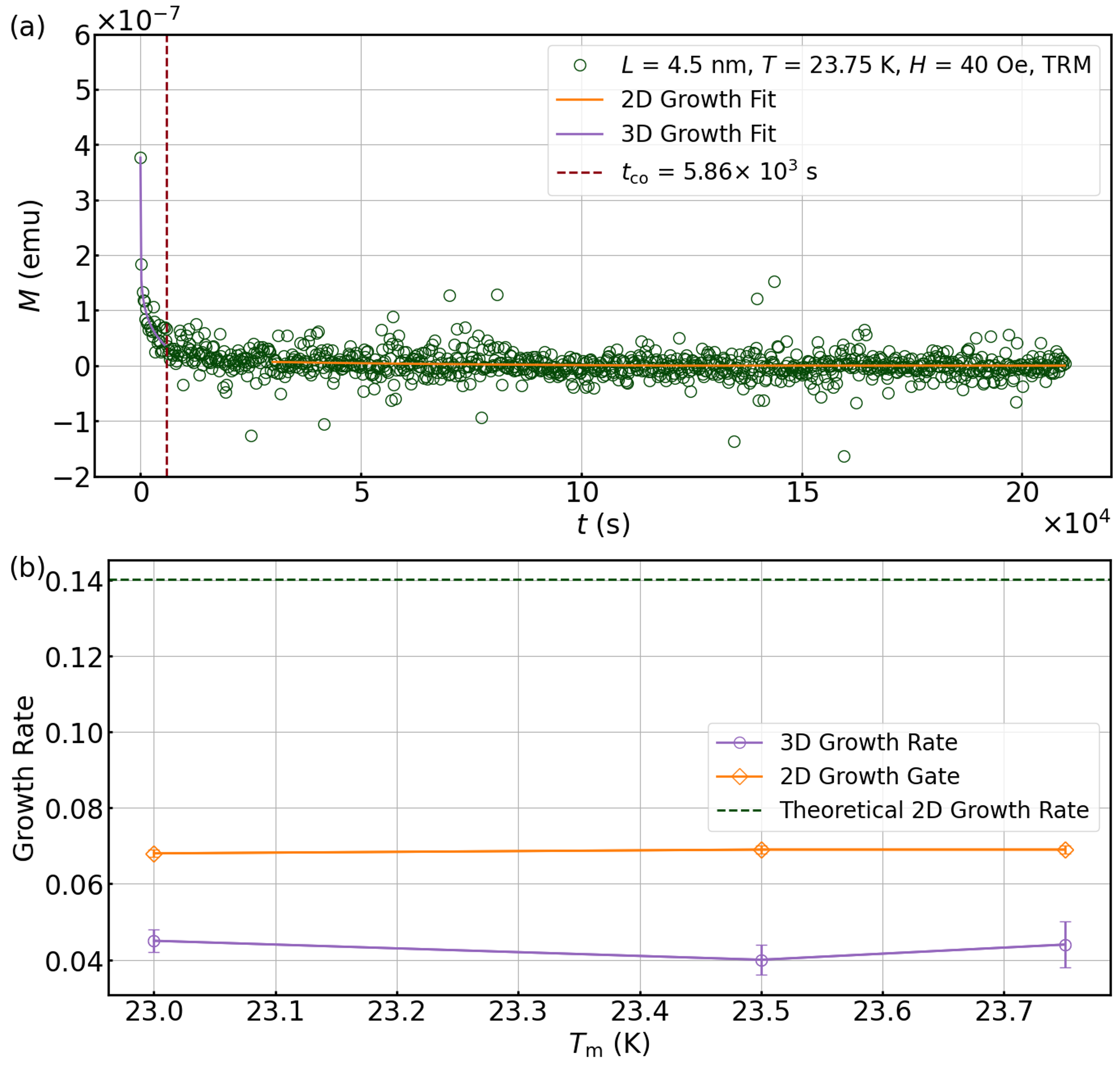}
    \caption{(a) Typical $M_\text{TRM}(t)$ at $T=23.75$ K and $H=40$ Oe with the fit in $D=3$ using Eq. \ref{eq:3D_fit_eq} and the fit in $D=2$ using Eq. \ref{eq:2D_fit_eq} overlay. The fit in $D=3$ starts at $t=0$ and ends at $t=t_\text{co}$; the fit in $D=2$ starts at $t=3\times10^4$ s and ends at our longest measurement time. (b) The temperature dependence of the growth rate $T/(T_\text{g}Z_c)$ in $D=3$ and $1/Z_c$ in $D=2$. The red dashed line shows the theoretical growth rate parameter $Z_c$ in $D=2$ with $Z_c \approx 7.14$ \cite{Fernandez:19b}.}
    \label{fig:fit_growth_rate}
\end{figure}

\noindent The $\alpha_f$-adjusted $M_\text{TRM}(t,T)$ shown in Fig. \ref{fig:2d_3d_dynamics}(a) demonstrates the temperature dependent dynamics in $D=3$ at short time scales. The different rates for the magnetic moment decay reflect the different growth rates of $\xi(t, T)$ at different temperatures, which result in different $t_\text{co}$. Fig. \ref{fig:2d_3d_dynamics}(b) shows the growth in $D=2$ using $\alpha_f$-adjusted $M_\text{TRM}(t-t_\text{co},T)$. The overlap of magnetic moment decays at $t>t_\text{co}$ is consistent with {\it temperature independent} dynamics in $D=2$ at long time scales. The deviation of decays at $t<t_\text{co}$ aligns with the {\it temperature dependent} dynamics in $D=3$.\\
\\
In our analysis, we adopt the power law growth rate, Eq. (3),  
to fit $M_\text{TRM}(t, T)$ obtained from Eq. (5) to extract the growth rate $T/(T_\text{g}Z_c)$ in $D=3$ and $1/Z_c$ in $D=2$. The exchange rate $1/\tau_0$ is $\sim k_\text{B}T_\text{g}/\hbar\approx6.9\times10^{12}$ s$^{-1}$. A cubic spline interpolation was used to reduce our data size to 1000 points. Typical power growth law fits for $D=3$ and $D=2$ are shown in Fig. \ref{fig:fit_growth_rate}(a).\\
\\
For growth in $D=3$, i.e., when $t\leq t_\text{co}$, we have
\begin{equation}
    M_\text{TRM}(t, T)=M_\text{TRM}(t=0, T)-m_\text{3D}[\xi_\text{3D}(t,T)]^D
    \label{eq:3D_fit_eq}
\end{equation}
where $m_\text{3D}$ is the amplitude to correlate the growth volume with the magnetic moment. When using Eq. \ref{eq:3D_fit_eq} to fit our experimental data at $t\leq t_\text{co}$, we fix $c_1=1.448$ and make $Z_c$ a fitting parameter.\\
\\
When $t> t_\text{co}$, the growth is in $D=2$, and we have
\begin{equation}
    M_\text{TRM}(t, T)=m_\text{2D}\frac{[\xi^\text{eq}_\text{2D}]^D-[\xi_\text{2D}(t_\text{adj},T)]^D}{[\xi^\text{eq}_\text{2D}]^D}
    \label{eq:2D_fit_eq}
\end{equation}
where $m_\text{2D}$ is the amplitude to correlate the growth volume with the magnetic moment. $t_\text{adj}$ is an adjusted time, assuming a hypothetical growth in $D=2$ starting from $\xi=0$. $t_\text{adj}$ is calculated using
\begin{equation}
    t_\text{adj}=t-t_\text{co}+t_\text{2D}
\end{equation}
where $t_\text{2D}$ is the time needed in the hypothetical growth in $D=2$ to reach $\xi=\mathcal{L}$, i.e.,
\begin{equation}
    \xi_\text{2D}(t_\text{2D})=a_0c_1(t_\text{2D}/\tau_0)^{1/Z_c^\text{2D}}=\mathcal{L}
\end{equation}
In the fit to our data at $t> t_\text{co}$, we assume $c_1=1$ as there is no experimental value for $c_1$ in $D=2$, and make $Z_c$ a fitting parameter. $t_\text{2D}$ is obtained from the fit result using an initial value, then the fit is iterated with the new $t_\text{2D}$ value until $t_\text{2D}$ converges.\\
\\
The temperature dependence of the growth rate in $D=3$ and $D=2$ is shown in Fig. \ref{fig:fit_growth_rate}(b). The {\it temperature independent} $D=2$ growth rate is faster than the {\it temperature dependent} $D=3$ growth rate. Although we find a growth rate in $D = 2$ greater than in $D = 3$, the ratio is less than that generated through the scaling approach \cite{Fernandez:19b}.\\

\begin{figure}[htbp]
    \centering
    \includegraphics[width=0.49\textwidth]{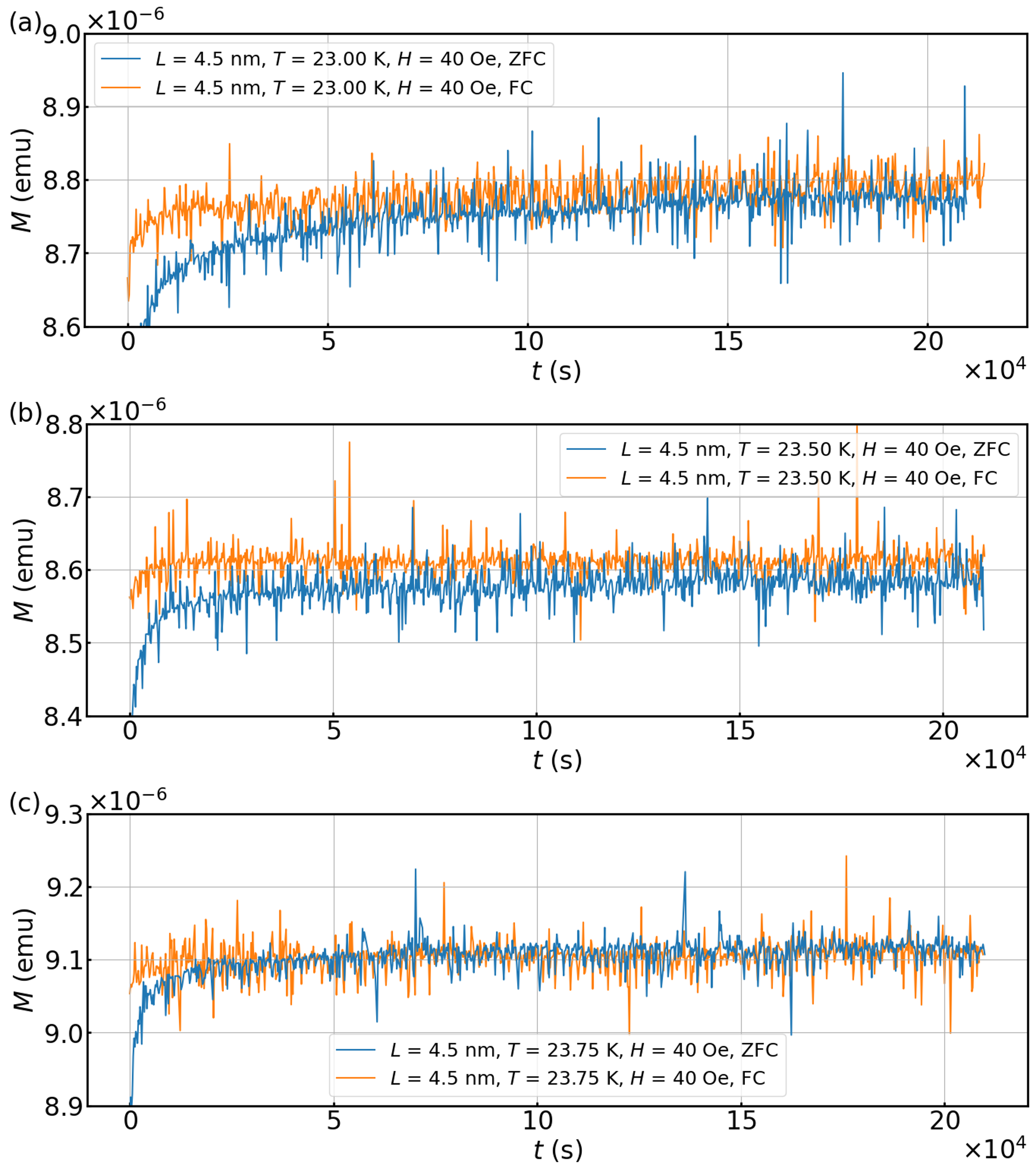}
    \caption{$M_\text{ZFC}(t,T)$ and $M_\text{FC}(t,T)$ at (a) $T_\text{m}=23.00$ K, (b) $T_\text{m}=23.50$ K, and (c) $T_\text{m}=23.75$ K for $L=4.5$ nm CuMn/Cu multilayer sample with $H=40$ Oe. The data of $M_\text{ZFC}(t,T)$ and $M_\text{FC}(t,T)$ is reduced to 1000 points using cubic spline interpolation to better demonstrate the differences at different temperatures.}
    \label{fig:zfc_fc_all_temps}
\end{figure}

\noindent Equilibrium is reached when the growth of $\xi$ completes ($\sim\xi^\text{eq}$), indicated by constant magnetic moments, and the difference between $M_\text{ZFC}(t,T)$ and $M_\text{FC}(t,T)$ diminishes. Fig. \ref{fig:zfc_fc_all_temps} shows the typical $M_\text{ZFC}(t,T)$ and $M_\text{FC}(t,T)$ data at our measurement temperatures. It is clearly shown in Fig. \ref{fig:zfc_fc_all_temps}(a) that equilibrium has not been reached at $T_\text{m}=23.00$ K: both $M_\text{ZFC}(t,T)$ and $M_\text{FC}(t,T)$ are still changing at long time scale. However, the system does reach equilibrium at $T_\text{m}=23.75$ K within our experimental time scale as shown in Fig. \ref{fig:zfc_fc_all_temps}(c): $M_\text{ZFC}(t,T)$ and $M_\text{FC}(t,T)$ stay constant at the same level at long time scale. The fit at $T_\text{m}=23.00$ K and $23.75$ K using unadjusted $M_\text{TRM}(t,T)$ yields the {\it temperature independent} growth rate $1/Z_c\approx0.069$ in $D=2$. \\
\\
Using the experimentally extracted $D=2$ growth rate, we can calculate the total time required to reach equilibrium $t^{\text{eq}}$:
\begin{equation}
    t^\text{eq}=t^\text{eq}_\text{2D}-t_\text{2D}+t_\text{co}
\end{equation}
where $t_\text{co}$ is the crossover time obtained previously, $t_\text{2D}$ and $t^\text{eq}_\text{2D}$ are the calculated times for $\xi_\text{2D}=0$ to reach $\mathcal{L}$ and $\xi_\text{2D}^\text{eq}$, respectively, using the power growth law with a growth rate $1/Z_c\approx0.069$ in $D=2$. The growth rate indicates that equilibrium should be reached at $T_\text{m}=23.50$ K within our experimental time scale, which aligns with our data shown in Fig. \ref{fig:zfc_fc_all_temps}(b). Although $M_\text{ZFC}(t,T)$ and $M_\text{FC}(t,T)$ stay constant at long time scale for $T_\text{m}=23.50$ K, there is still a finite difference between them. This prompts us to use $\alpha_f$-adjusted $M_\text{TRM}(t,T)$ at $T_\text{m}=23.50$ K to obtain the results shown in Fig. \ref{fig:fit_growth_rate}(b). Table \ref{tab:eq_alphaf} lists $t^{\text{eq}}$ at all $T_\text{m}$, and $\alpha_f$ is used to adjust $M_\text{TRM}(t,T)$ for our measurements that reach equilibrium.\\
\begin{table}[htbp]
    \centering
    \caption{The total time to reach equilibrium $t^{\text{eq}}$, whether the equilibrium is reached in our experimental time scale, the scaling factor at long time scale $\alpha_f$, and whether $\alpha_f$ is used to adjust $M_\text{TRM}(t)$ at different measurement temperatures $T_\text{m}$.}
    \begin{ruledtabular}
        \begin{tabular}{cccccccc}
       $T_\text{m}$(K) & $t^{\text{eq}}$ (s) & Equilibrium & $\alpha_f$ & Adjusted \\
        \hline
        23.00 & $5.32\times10^5$ & N & 0.997 & N \\
        23.50 & $1.79\times10^5$ & Y & 0.997 & Y \\
        23.75 & $1.05\times10^5$ & Y & 1.000 & Y \\
        \end{tabular}
    \end{ruledtabular}
    \label{tab:eq_alphaf}
\end{table}

\noindent To summarize, we have been able to probe the growth rate of the correlation function for spin glasses in $D = 2$.  We have shown that it is faster than in $D = 3$, but less fast than predicted from a scaling approach.  Further, we have demonstrated that we can achieve equilibrium for the correlation function in $D = 2$ at sufficiently long measurement times.  Both of these achievements are unique, and a consequence of using thin film multilayers to reach a stable $D = 2$ spin glass.  Our approach can be generalized to other systems where the spatial dimension is less than the lower critical dimension, generating new opportunities for investigations.\\
\\
This work was supported by the U.S. Department of Energy, Office of Science, Basic Energy Sciences, Division of Materials
Science and Engineering, under Award No. DE-SC0013599.
H.L. acknowledges the partial support by the National Science
Foundation through the Center for Dynamics and Control of
Materials, an NSF MRSEC under Cooperative Agreement No.
DMR-2308817.

\makeatletter
\def\bibsection{%
  \par
  \baselineskip26\p@
  \bib@device{\linewidth}{82\p@}%
  \nobreak\@nobreaktrue
  \addvspace{19\p@}%
  \par
}
\makeatother
\bibliography{citations}

@article{Bray:86,
  title = {Lower Critical Dimension of Metallic Vector Spin-Glasses},
  author = {Bray, A. J. and Moore, M. A. and Young, A. P.},
  journal = {Phys. Rev. Lett.},
  volume = {56},
  issue = {24},
  pages = {2641--2644},
  numpages = {0},
  year = {1986},
  month = {Jun},
  publisher = {American Physical Society},
  doi = {10.1103/PhysRevLett.56.2641},
  url = {https://link.aps.org/doi/10.1103/PhysRevLett.56.2641}
}

@article{Boettcher:05,
  title = {Stiffness of the {Edwards-Anderson} Model in all Dimensions},
  author = {Boettcher, Stefan},
  journal = {Phys. Rev. Lett.},
  volume = {95},
  issue = {19},
  pages = {197205},
  numpages = {4},
  year = {2005},
  month = {Nov},
  publisher = {American Physical Society},
  doi = {10.1103/PhysRevLett.95.197205},
  url = {https://link.aps.org/doi/10.1103/PhysRevLett.95.197205}
}

@article{Franz:94,
    author = {S. Franz and G. Parisi and M. Virasoro},
    title = {Interfaces and louver critical dimension in a spin glass model},
    journal = {J. Phys (France)},
    volume = {4},
    page = {1657},
    year = {1994}
}

@article{Maiorano:18,
    author = {A. Maiorano and G. Parisi},
    title = {Support for the value 5/2 for the spin glass lower critical dimension at zero magnetic field},
    journal = {Proc. Natl. Acad, Sci. USA},
    Volume = {115},
    Page = {5129},
    year = {2018}
}

@article{Dahlberg:25,
  title = {Spin-glass dynamics: Experiment, theory, and simulation},
  author = {Dahlberg, E. D. and Gonz\'alez-Adalid Pemart\'{\i}n, I. and Marinari, E. and Parisi, G. and Ricci-Tersenghi, F. and Martin-Mayor, V. and Moreno-Gordo, J. and Orbach, R. L. and Paga, I. and Ruiz-Lorenzo, J. J. and Yllanes, D.},
  journal = {Rev. Mod. Phys.},
  volume = {97},
  issue = {4},
  pages = {045005},
  numpages = {54},
  year = {2025},
  month = {Dec},
  publisher = {American Physical Society},
  doi = {10.1103/ctp2-zwyr},
  url = {https://link.aps.org/doi/10.1103/ctp2-zwyr}
}

@article{Guchhait:15,
  title = {Spin glass dynamics at the mesoscale},
  author = {Guchhait, Samaresh and Kenning, Gregory G. and Orbach, Raymond L. and Rodriguez, Gilberto F.},
  journal = {Phys. Rev. B},
  volume = {91},
  issue = {1},
  pages = {014434},
  numpages = {10},
  year = {2015},
  month = {Jan},
  publisher = {American Physical Society},
  doi = {10.1103/PhysRevB.91.014434},
  url = {https://link.aps.org/doi/10.1103/PhysRevB.91.014434}
}

@article{Zhai:17a,
  title = {Glassy dynamics in {CuMn} thin-film multilayers},
  author = {Zhai, Qiang and Harrison, David C. and Tennant, Daniel and Dahlberg, E. Dan and Kenning, Gregory G. and Orbach, Raymond L.},
  journal = {Phys. Rev. B},
  volume = {95},
  issue = {5},
  pages = {054304},
  numpages = {8},
  year = {2017},
  month = {Feb},
  publisher = {American Physical Society},
  doi = {10.1103/PhysRevB.95.054304},
  url = {https://link.aps.org/doi/10.1103/PhysRevB.95.054304}
}

@article{Zhai:17b,
  title = {Effect of magnetic fields on spin glass dynamics},
  author = {Zhai, Qiang and Harrison, David C. and Orbach, Raymond L.},
  journal = {Phys. Rev. B},
  volume = {96},
  issue = {5},
  pages = {054408},
  numpages = {6},
  year = {2017},
  month = {Aug},
  publisher = {American Physical Society},
  doi = {10.1103/PhysRevB.96.054408},
  url = {https://link.aps.org/doi/10.1103/PhysRevB.96.054408}
}

@article{Kenning:90,
  title = {Finite-size effects in {Cu-Mn} spin glasses},
  author = {Kenning, G. G. and Bass, Jack and Pratt, W. P. and Leslie-Pelecky, D. and Hoines, Lilian and Leach, W. and Wilson, M. L. and Stubi, R. and Cowen, J. A.},
  journal = {Phys. Rev. B},
  volume = {42},
  issue = {4},
  pages = {2393--2415},
  numpages = {0},
  year = {1990},
  month = {Aug},
  publisher = {American Physical Society},
  doi = {10.1103/PhysRevB.42.2393},
  url = {https://link.aps.org/doi/10.1103/PhysRevB.42.2393}
}

@article{Zhai:24,
    author = {Q. Zhai and R.L. Orbach},
    title = {Toward Understanding the dimensional crossover of canonical spin-glass films},
    journal = {Frontiers in Physics},
    year = {2024},
    volume = {12},
    page = {1488275}
}

@article{Fernandez:19a,
  title = {Dimensional crossover in the aging dynamics of spin glasses in a film geometry},
  author = {Fernandez, L. A. and Marinari, E. and Martin-Mayor, V. and Paga, I. and Ruiz-Lorenzo, J. J.},
  journal = {Phys. Rev. B},
  volume = {100},
  issue = {18},
  pages = {184412},
  numpages = {8},
  year = {2019},
  month = {Nov},
  publisher = {American Physical Society},
  doi = {10.1103/PhysRevB.100.184412},
  url = {https://link.aps.org/doi/10.1103/PhysRevB.100.184412}
}

@article{Fernandez:19b,
    author = {L.A. Fernandez and E. Marinari and V. Martin-Mayor and G. Parisi and J.J. Ruiz-Lorenzo},
    title = {An experimental-oriented analysis of {2D} spin-glass dynamics: a twelve time-decades scaling study},
    journal = {J. Phys. A: Math. Theor.},
    volume = {52},
    year = {2019},
    publisher = {IOP Publishing},
    pages = {224002},
    
}

@article{PhysRevLett.54.150,
  title = {Evidence for Multiple Mechanisms Contributing to the Transition Temperature in Metallic Spin-Glasses},
  author = {Vier, D. C. and Schultz, S.},
  journal = {Phys. Rev. Lett.},
  volume = {54},
  issue = {2},
  pages = {150--153},
  numpages = {0},
  year = {1985},
  month = {Jan},
  publisher = {American Physical Society},
  doi = {10.1103/PhysRevLett.54.150}
}

@article{PhysRevB.33.4803,
  title = {Saturation of {Ruderman-Kittel-Kasuya-Yosida} interaction damping in high-resistivity spin glasses},
  author = {Larsen, Ulf},
  journal = {Phys. Rev. B},
  volume = {33},
  issue = {7},
  pages = {4803--4808},
  numpages = {0},
  year = {1986},
  month = {Apr},
  publisher = {American Physical Society},
  doi = {10.1103/PhysRevB.33.4803},
  url = {https://link.aps.org/doi/10.1103/PhysRevB.33.4803}
}

@inbook{doi:10.1142/9789812819437_0001,
	author = {Per Nordblad and Peter Svedlindh},
	booktitle = {Spin Glasses and Random Fields},
	doi = {10.1142/9789812819437_0001},
	pages = {1-27},
    year = {1997},
    publisher = {World Scientific Publishing},
	title = {EXPERIMENTS ON SPIN GLASSES}}

@article{Baity:18,
  title = {Aging Rate of Spin Glasses from Simulations Matches Experiments},
  author = {Baity-Jesi, M. and Calore, E. and Cruz, A. and Fernandez, L. A. and Gil-Narvion, J. M. and Gordillo-Guerrero, A. and I\~niguez, D. and Maiorano, A. and Marinari, E. and Martin-Mayor, V. and Moreno-Gordo, J. and Mu\~noz-Sudupe, A. and Navarro, D. and Parisi, G. and Perez-Gaviro, S. and Ricci-Tersenghi, F. and Ruiz-Lorenzo, J. J. and Schifano, S. F. and Seoane, B. and Tarancon, A. and Tripiccione, R. and Yllanes, D.},
  collaboration = {Janus Collaboration},
  journal = {Phys. Rev. Lett.},
  volume = {120},
  issue = {26},
  pages = {267203},
  numpages = {6},
  year = {2018},
  month = {Jun},
  publisher = {American Physical Society},
  doi = {10.1103/PhysRevLett.120.267203},
  url = {https://link.aps.org/doi/10.1103/PhysRevLett.120.267203}
}

@misc{Young:14,
       author = {A. P. Young},
       year = {2014},
       title = {private communication},  
}

@misc{Victor:16,
        author = {V. Martin-Mayor},
        year = {2016},
        title = {private communication},
}

@article{Dekker:88a,
    author = {C. Dekker and A.F.M. Arts and H.W. de Wijn},
    title = {{Rb$_2$Cu$_{1-x}$Co$_x$F$_4$}, a two-dimensional Ising spin glass},
    journal = {Journal of Applied Physics},
    year = {1988},
    page = {4334},
    volume = {63},
    number = {8}
}

@article{Dekker:88b,
  title = {Activated Dynamics in the Two-Dimensional Ising Spin-Glass {${\mathrm{Rb}}_{2}{\mathrm{Cu}}_{1\ensuremath{-}x}{\mathrm{Co}}_{x}{\mathrm{F}}_{4}$}},
  author = {Dekker, C. and Arts, A. F. M. and de Wijn, H. W. and van Duyneveldt, A. J. and Mydosh, J. A.},
  journal = {Phys. Rev. Lett.},
  volume = {61},
  issue = {15},
  pages = {1780--1783},
  numpages = {0},
  year = {1988},
  month = {Oct},
  publisher = {American Physical Society},
  doi = {10.1103/PhysRevLett.61.1780},
  url = {https://link.aps.org/doi/10.1103/PhysRevLett.61.1780}
}

@article{Dekker:88c,
  title = {Static critical behavior of the two-dimensional Ising spin glass {${\mathrm{Rb}}_{2}$${\mathrm{Cu}}_{1\mathrm{\ensuremath{-}}\mathrm{x}}$${\mathrm{Co}}_{\mathrm{x}}$${\mathrm{F}}_{4}$}},
  author = {Dekker, C. and Arts, A. F. M. and de Wijn, H. W.},
  journal = {Phys. Rev. B},
  volume = {38},
  issue = {13},
  pages = {8985--8991},
  numpages = {0},
  year = {1988},
  month = {Nov},
  publisher = {American Physical Society},
  doi = {10.1103/PhysRevB.38.8985},
  url = {https://link.aps.org/doi/10.1103/PhysRevB.38.8985}
}

@article{Dekker:89,
  title = {Activated dynamics in a two-dimensional Ising spin glass: {${\mathrm{Rb}}_{2}$${\mathrm{Cu}}_{1\mathrm{\ensuremath{-}}\mathrm{x}}$${\mathrm{Co}}_{\mathrm{x}}$${\mathrm{F}}_{4}$}},
  author = {Dekker, C. and Arts, A. F. M. and de Wijn, H. W. and van Duyneveldt, A. J. and Mydosh, J. A.},
  journal = {Phys. Rev. B},
  volume = {40},
  issue = {16},
  pages = {11243--11251},
  numpages = {0},
  year = {1989},
  month = {Dec},
  publisher = {American Physical Society},
  doi = {10.1103/PhysRevB.40.11243},
  url = {https://link.aps.org/doi/10.1103/PhysRevB.40.11243}
}

\end{document}